\documentclass[useAMS,usenatbib,usegraphicx]{mn2e}

\usepackage{amsmath}
\usepackage{amsfonts}
\usepackage{amssymb}
\usepackage{color}

\newcommand{\aap}{A\&A} 
 
\newcommand{\mnras}{MNRAS} 
\newcommand{\apj}{ApJ} 
\newcommand{\araa}{ARA\&A}

\newcommand{\aj}{AJ}

\newcommand{\pasj}{PASJ}

\newcommand{\dd}{\mbox{d}}
\newcommand{\ii}{_{\mbox{\tiny init.}}}
\newcommand{\maxx}{_{\mbox{\tiny max}}}
\newcommand{\rapo}{_{\mbox{\tiny apo.}}}
\newcommand{\rperi}{_{\mbox{\tiny peri.}}}
\newcommand{\select}{_{\mbox{\tiny select}}} 
\newcommand{\df}{_{\mbox{\tiny DF}}} 

\title[Probe Galactic potential via action-based DF]{Constraining the Galactic potential via action-based distribution functions for mono-abundance stellar populations}
\author[Y.-S. Ting et al.]{Yuan-Sen Ting,$^{1,2}$\thanks{E-mail: yuan-sen.ting@cfa.harvard.edu} Hans-Walter Rix,$^{2}$ Jo Bovy$^{3\mbox{\normalsize \thanks{Hubble fellow}}}$ and Glenn van de Ven$^{2}$ \\
$^{1}$Harvard--Smithsonian Center for Astrophysics, 60 Garden Street, Cambridge, MA 02139, USA\\
$^{2}$Max Planck Institute for Astronomy, K\"onigstuhl 17, D-69117 Heidelberg, Germany\\
$^{3}$Institute for Advanced Study, Einstein Drive, Princeton, NJ 08540, USA
}

\begin{document}

\date{Accepted 2013 June 11. Received 2013 June 10; in original form 2012 November 29}

\maketitle

\label{firstpage}

%
%
%
%
%
%
\begin{abstract}
We present a rigorous and practical way of constraining the Galactic potential based on the phase-space information for many individual stars. Such an approach is needed to dynamically model the data from ongoing spectroscopic surveys of the Galaxy and in the future {\em Gaia}. This approach describes the orbit distribution of stars by a family of parametrized distribution function (DF) proposed by McMillan and Binney, which are based on actions. We find that these parametrized DFs are flexible enough to capture well the observed phase-space distributions of individual abundance-selected Galactic subpopulations of stars (`mono-abundance populations') for a disc-like gravitational potential, which enables independent dynamical constraints from each of the Galactic mono-abundance populations. 

We lay out a statistically rigorous way to constrain the Galactic potential parameters by constructing the joint likelihood of potential and DF parameters, and subsequently marginalizing over the DF parameters. This approach explicitly incorporates the spatial selection function inherent to all Galactic surveys, and can account for the uncertainties of the individual position--velocity observations. 

On that basis, we study the precision of the parameters of the Galactic potential that can be reached with various sample sizes and realistic spatial selection functions. By creating mock samples from the DF, we show that, even under a restrictive and realistic spatial selection function, given a two-parameter gravitational potential, one can recover the true potential parameters to a few per cent with sample sizes of a few thousands. The assumptions of axisymmetry, of DFs that are smooth in the actions and of no time variation remain important limitations in our current study.
\end{abstract}
 

%
%
%
%
%
%
\section[]{Introduction}\label{sec:introduction}
Constraining the Galactic potential optimally has been a very difficult yet important aspect of studies of the Milky Way and the Local Group. Understanding the various components of the Galactic potential, for instance separating the potential contributions from the baryonic disc and that from the dark matter halo, is fundamental to understanding the history and formation of our own Galaxy. Furthermore, understanding the dark halo potential near the Sun is a crucial step to pin down the density and thus the scattering cross-section of the dark matter, which in turn is crucial input for interpreting any dark matter annihilation signal at the centre of Milky Way \citep[e.g.][]{su12}.

In the past few decades, constraining the Galactic potential has mostly relied on the Jeans equation \citep[for a summary, see][]{bin08} despite the known problems of this approach. For instance, one only keeps the velocity dispersion moments up to second order in the Jeans equation. This predicts a Gaussian ellipsoid velocity distribution that does not capture the observed and expected asymmetries in the angular velocity $v_\phi$ \citep[e.g.][]{fuc09}. 

One of the ways to alleviate these problems is through explicitly modelling the stellar population distribution function (DF), either via integrals of motions or actions instead of just taking position--velocity moments, as is the case of the Jeans equation \citep*{deh99,bin10,bin11b,mcm12,sol12}. For such modelling, it is crucial that relatively simple analytic DFs, e.g. based on the actions, are reasonably good approximations to the actual orbit distributions for at least subpopulations of stars in the Galactic disc. How well such DF approximations work in practice is not yet clear. To explore this question, we consider in particular stellar subpopulations of a particular age or abundance pattern \citep{fre02,tin12}. 

Recently, \citet*{bov12a} and \citet*{bov12b,bov12c} (hereafter refer all of these references as BO12) showed, using G-dwarfs from the Sloan Extension for Galactic Understanding and Exploration (SEGUE) survey \citep{yan09}, that mono-abundance, stars that have nearly the same [Fe/H] and [$\alpha$/Fe], populations have very simple $({\boldsymbol x},{\boldsymbol v})$ phase-space distribution properties: spatial density distributions that vary exponentially, both in the radial and vertical directions; furthermore, each mono-abundance population shows isothermal velocity dispersion in the vertical direction and exponentially in the radial direction \citep{bov12b}. This simplicity in $({\boldsymbol x},{\boldsymbol v})$ space gives hope that one could describe mono-abundance populations with simple action-based DFs which is one of the purposes of this study. 

Since stars are the most abundant tracers of the Galactic potential, simple action-based DFs will naturally provide a statistical and rigorous method to constrain the Galactic potential because the conversion of position--velocity variable (configuration space) to the action--angle variable depends on the parameters of the potential (see \citealt{bin10,bin11a,bin12,bin11b,mcm11,mcm12} for generic ground work on the topic). The main idea is to determine the likelihood of the observational data, given a joint set of parameters for both the DF and the gravitational potential; subsequent marginalization over the DF parameters provides then a rigorously derived constraint on the potential. Such an approach appears to be a precondition to fully exploit the dynamical information content of large Galactic surveys that will provide us spatial and velocity distribution, along with elemental abundances, including the Apache Point Observatory Galactic Evolution Experiment \citep{eis11}, Galactic Archaeology survey with HERMES\footnote{High Efficiency and Resolution Multi-Element Spectrograph} \citep[GALAH;][]{fre10,fre12} and European Southern Observatory-{\em Gaia} \citep{gil12}.

In the context of dynamically modelling the Galactic disc with action-based DFs, the goals of the paper are three-fold: explore how well `mono-abundance populations' can be approximated by simple action-based analytic DFs; lay out a formalism that provides constraints from a set of discrete stellar positions and velocities on the gravitational potential, after marginalizing over the DF; forecast what constraints on the {\em shape} of the potential we can expect given existing sample sizes.

This paper is organized as follows: in Section\,\ref{sec:DF}, we discuss the choice of parametrization of the DF and summarize the adiabatic approximation that we assume to calculate the action variable efficiently. In Section\,\ref{sec:DF-match}, we will show that our choice of DF exhibits all the basic phase-space distribution properties of mono-abundance stars, with the only caveat that disc potential has to be included in order for the vertical spatial profile to fit. In Section\,\ref{sec:constrain_potential}, we show that by studying the likelihood of the DF--potential parameters, one can recover the potential parameters, even under a restrictive and realistic spatial selection function. 

Throughout this study, we denote the cylindrical coordinate to be ${\boldsymbol x}\equiv(R,\phi,z)$. We assume that velocities are measured in the inertial Galactocentric frame with axisymmetric potential, $\Phi=\Phi(R,z)$. In practice, the conversion from the heliocentric frame to the Galactocentric frame requires the knowledge of the solar motion. From the study of Sgr\,A*, the solar motion is now accurate to a few $\mbox{km}\,\mbox{s}^{-1}$ \citep[e.g.][]{rei04} and the solar radius $R_{\mbox{\sun}}$ is accurate to $\sim0.3-0.4$\,kpc \citep[e.g.][]{ghe08,gil09}. We will also discuss the uncertainty in estimating the potential parameters due to the uncertainty of this conversion in Section\,\ref{sec:constrain_potential}.

%
%
%
%
%
%
\section[]{Choice of the DF}\label{sec:DF}
We follow the DF advocated by \citet{bin10} and \citet{mcm12} closely, but note that we rearrange some of the terms in the DF to facilitate physical interpretation of the DF. For circular orbits in the equatorial Galactic plane, the DF has to be of the form
\begin{eqnarray}
f(J_R,L_z,J_z)=\widetilde{f}(L_z)\delta(J_R-0)\delta(J_z-0),
\end{eqnarray}

\noindent
where $\widetilde{f}(L_z)$ determines the angular momentum distribution for an infinitely cold planar disc, $J_R$ and $J_z$ are the radial and vertical action, respectively. 

In the cold disc limit, the circular radius $R_c$ coincides with the observed radius, where
\begin{eqnarray}
L_z=R_c V_c(R_c)
\end{eqnarray}

\noindent
and $V_c$ is the circular velocity,
\begin{eqnarray}
V_c=\sqrt{R \frac{\upartial\Phi}{\upartial R}}.
\end{eqnarray}

Thus, we have for an exponential disc, up to a multiplicative normalization constant,
\begin{eqnarray} 
\widetilde{f}(L_z)\dd L_z=\exp\bigg(-\frac{R_c}{h_R}\bigg)R_c\dd R_c.
\end{eqnarray}

\noindent
The DF parameter $h_R$ determines the radial spatial distribution. In a more general setting, we relax the $J_R=J_z=0$ constraint and propose the DF to be
\begin{eqnarray}
f(J_R,L_z,J_z)=\widetilde{f}(L_z)\bigg[\frac{\kappa(L_z)}{C_R(L_z)} \exp\bigg(-\frac{\kappa (L_z) J_R}{C_R(L_z)}\bigg)\bigg]\notag\\
\bigg[\frac{\nu(L_z)}{D_R(L_z)}\exp\bigg(-\frac{\nu (L_z) J_z}{D_R(L_z)}\bigg)\bigg],
\label{eq:DF}
\end{eqnarray}

\noindent
where
\begin{eqnarray}
C_R(L_z)=\sigma^2_{R}\exp\bigg(-\frac{R_c(L_z)}{h_\sigma}\bigg)
\end{eqnarray}
\begin{eqnarray}
D_R(L_z)=\sigma^2_{z}\exp\bigg(-\frac{R_c(L_z)}{h_\sigma}\bigg),
\end{eqnarray}

\noindent
with $R_c$ the circular radius given angular momentum $L_z$, assuming equatorial in-plane movement; $\kappa(L_z)$ and $\nu(L_z)$ are the radial and vertical frequency, respectively, under epicycle approximation. The DF parameter $h_\sigma$ determines the radial exponential decay of the velocity dispersion, whereas $\sigma_R^2$ and $\sigma_z^2$ control the total radial and vertical dispersion, respectively.\footnote{However, it must be borne in mind that the DF parameters $\sigma_R^2$ and $\sigma_z^2$ are {\em different} from the velocity dispersions obtained as moments of the DF, which correspond to the observable dispersion. These parameters are named as such because they govern the velocity dispersions. They differ in their value quite significantly from the velocity dispersions in the $({\boldsymbol x},{\boldsymbol v})$ space.}

The $\kappa$ and $\nu$ terms are necessary \citep[e.g.][]{bin10} for two reasons.
\begin{enumerate}
\item A finite increment in energy $E$ can lead to an infinite increment of actions in the unbound case. One would love to have a term that will tend to zero at large $L_z$ and couple this term with $J_R$. One of the possible choices is the radial frequency $\kappa(R_c(L_z))$. At large $R_c$, the effective potential is less concave at the minimum point and therefore $\kappa(R_c(L_z))$ tends to zero. Similarly for the vertical frequency.
\item Qualitatively, in $R_c$ region where $\kappa$ changes more drastically, the increment of $J_R$ is also more drastic, and therefore the scalelength of $J_R$ has to decrease in proportion to compensate for this effect. Similarly for the vertical oscillation. 
\end{enumerate}

%
%
%
%
%
%
\subsection[]{Adiabatic approximation}\label{subsec:adiabatic}
Importantly, we only measure $(R,z,v_R,v_\phi,v_z)$ but not the actions. In order to study an action-based DF, we employ the adiabatic approximation in order to calculate the radial and vertical actions efficiently. 
\begin{figure}
\begin{center}
\includegraphics[width=3.2in]{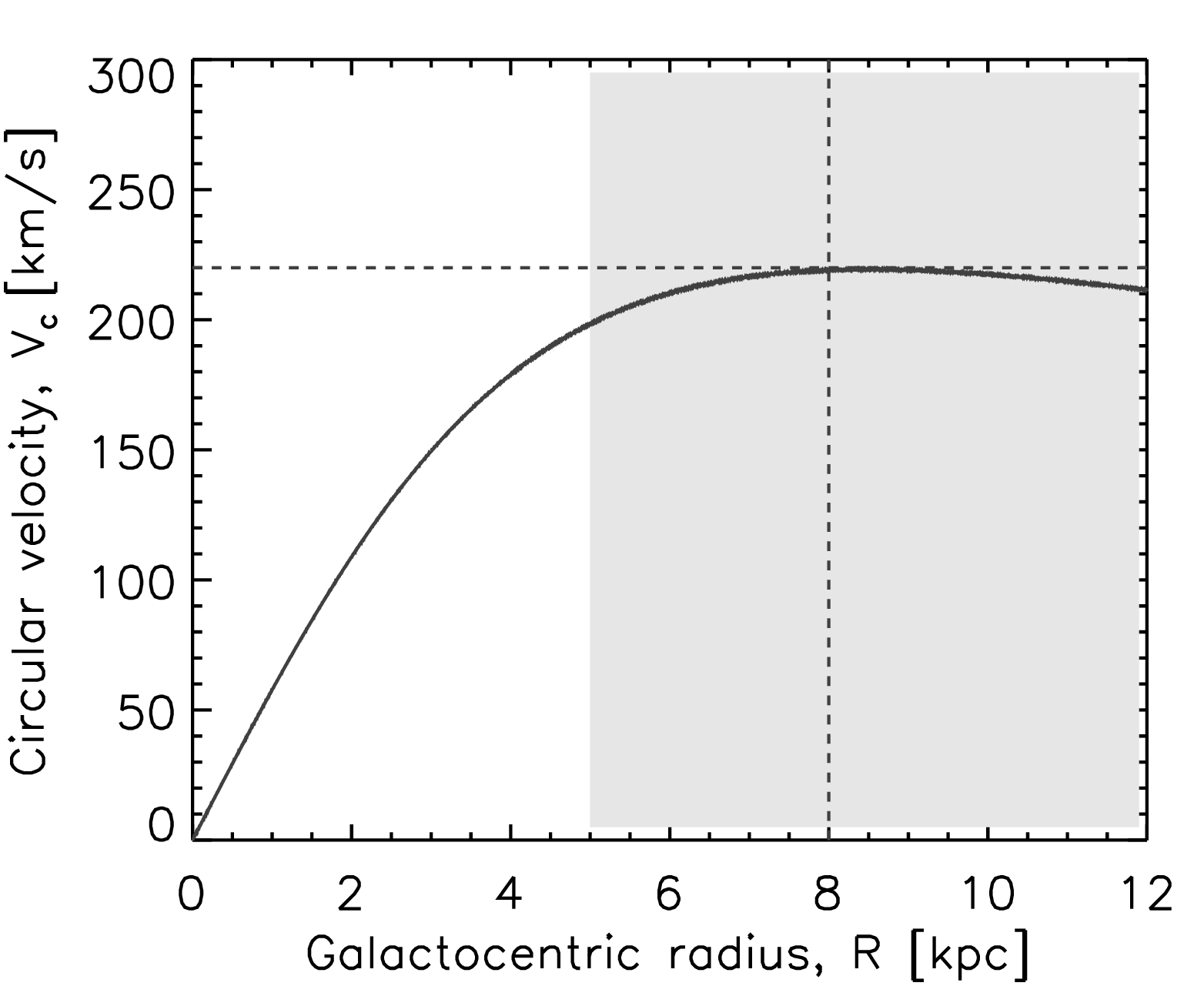}
\includegraphics[width=3.3in]{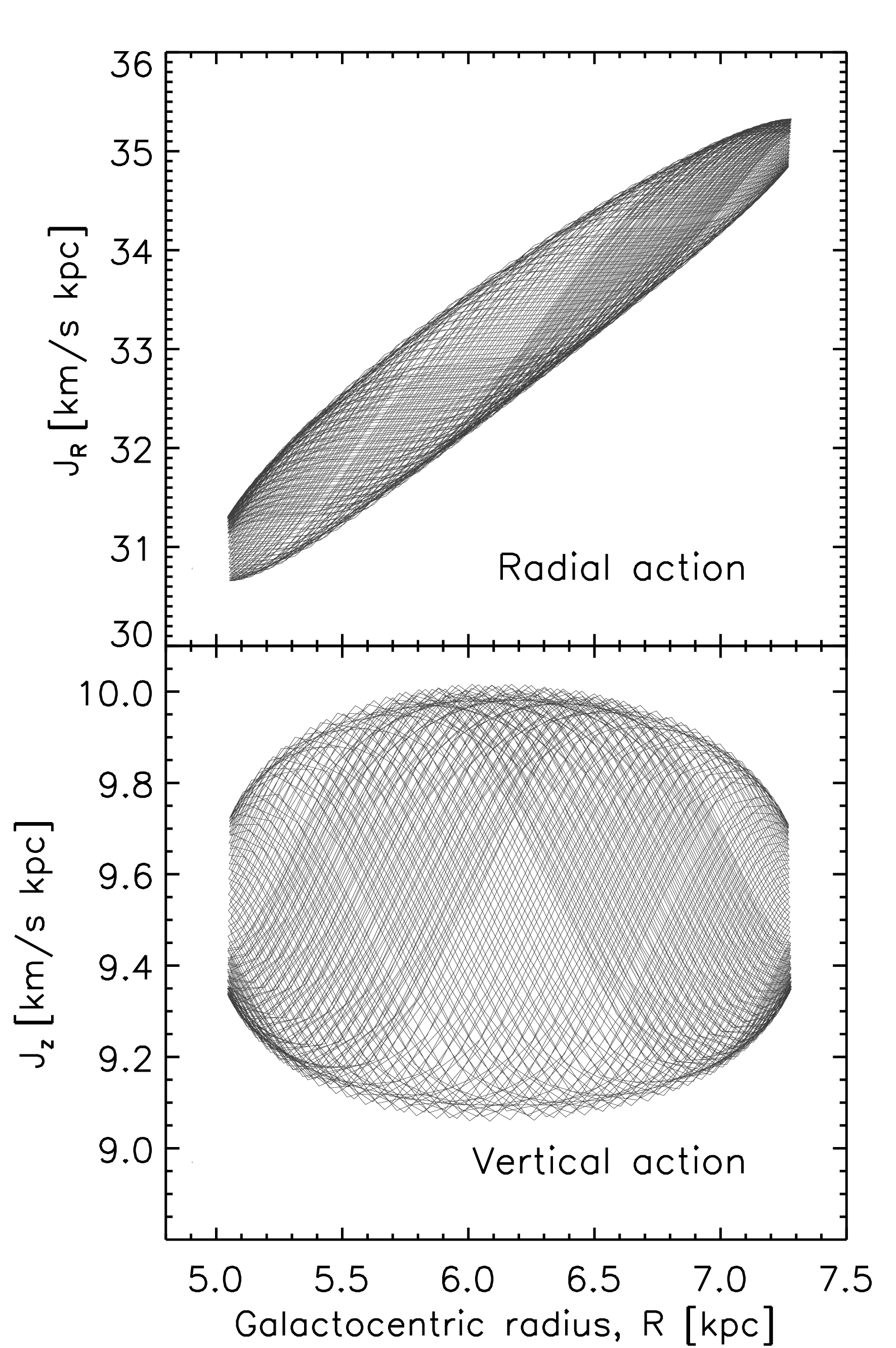}
\caption{Approximate action calculation in a Miyamoto--Nagai potential: the top panel shows the circular velocity of a Miyamoto--Nagai potential with $GM=7.5\times10^5\,\mbox{km}^2\,\mbox{s}^{-2}\,\mbox{kpc}^{-1}$, $a=5$\,kpc, $b=1$\,kpc, the vertical dashed line marks approximately the solar Galactocentric radius, the horizontal dashed line indicates the common accepted circular velocity $V_c=220\,\mbox{km}\,\mbox{s}^{-1}$ at the solar radius and the shaded region shows the region where the circular velocity is approximately constant in this choice of potential; the middle and bottom panels show the variation of actions calculated using the adiabatic approximation in the course of trajectory. We plot the case of $R\ii=6$ in Table\,\ref{table:adiabatic}. The results show that actions calculated with the adiabatic approximation are almost conserved along the orbit with less than $5$\,per cent variation. The orbit was integrated about 10 Gyr.}\label{fig:adiabatic}
\end{center}
\end{figure}

It has been shown that the vertical action $J_z$ calculated in a fixed radius $R$ approximation is almost conserved in the axisymmetric system \citep{bin11b}, and in discs including bar/spiral structure perturbation and radial migration \citep{min12,sol12}. More precisely,
\begin{eqnarray} 
J_z\simeq\oint\dd z\;\dot{z}(z|R).
\end{eqnarray}

This reduces the calculation to one-dimensional integration that can be numerically calculated easily. Furthermore, since $\Phi(z|R)=\Phi(-z|R)$, to calculate the approximated vertical action, we have
\begin{eqnarray}
J_z=\frac{2}{\upi}\int^{z\maxx}_0 dz\;\sqrt{\dot{z}\ii^2+2\bigg(\Phi(z\ii|R\ii)- \Phi(z|R\ii)\bigg)},
\end{eqnarray}

\noindent
where the subscript `init.' stands for the observed value in practice. 

Similarly, for the radial action, we can consider the radial component at the equatorial plane, ignoring the vertical motion. Under this assumption, 
\begin{eqnarray} 
J_R=\oint\dd R\;\dot{R}(R|z=0).
\end{eqnarray}
\begin{table*}
\begin{center}
\caption{Variation of the action variable under the adiabatic approximation with slight perturbation to the circular orbit under the Miyamoto--Nagai potential that gives a flat rotation curve.\label{table:adiabatic}}
\vspace{0.2cm}
\begin{tabular}{ccccccc}
\hline \hline \\[-0.3cm]
$R\ii$ & $z\ii$ & $L_z$ & $\dot{R}\ii$ & $\dot{z}\ii$ & $\Delta_{J_R}$ & $\Delta_{J_z}$ \\
(kpc)  & (kpc)  & $(\mbox{km}\,\mbox{s}^{-1}\,\mbox{kpc})$ & $(\mbox{km}\,\mbox{s}^{-1})$ & $(\mbox{km}\,\mbox{s}^{-1})$ & (per cent) & (per cent) \\
\hline \\[-0.3cm]
3      & 0      & 450   & 90           & 60           & 4              & 3 \\
6      & 0      & 1250  & 60           & 40           & 5              & 3 \\
9      & 0      & 1970  & 30           & 20           & 4              & 2 \\
\hline
\end{tabular}
\end{center}
\end{table*}

\noindent
We deduce
\begin{eqnarray}
J_R=\frac{1}{\upi}\int^{R\rapo}_{R\rperi}\dd R\notag
\end{eqnarray}
\begin{eqnarray}
\qquad\sqrt{\dot{R}^2\ii+\frac{ L_z^2}{R^2\ii}-\frac{L^2_z}{R^2}+2\bigg(\Phi(R\ii,0)-\Phi(R,0)\bigg)},
\end{eqnarray}

\noindent
where $R\rapo$ and $R\rperi$ are the apo- and pericentric radius, respectively. 

\subsection[]{Variation of actions under the adiabatic approximation}
Fig.\,\ref{fig:adiabatic} and Table\,\ref{table:adiabatic} illustrate the variation of the approximated actions in the course of a trajectory. We assume a Miyamoto--Nagai potential \citep{miy75},
\begin{eqnarray}
\label{eq:miyamoto_nagai}
\Phi(R,z)=-\frac{GM}{\sqrt{R^2+(a+\sqrt{z^2+b^2})^2}}
\end{eqnarray}

\noindent
with parameters $GM=7.5\times10^5\,\mbox{km}^2\,\mbox{s}^{-2}\,\mbox{kpc}^{-1}$, $a=5$\,kpc, $b=1$\,kpc. We choose these parameters because we have almost a flat rotation curve at the Galactocentric radius $5<R<12$\,kpc and circular velocity $V_c(R_{\mbox{\sun}})\simeq220\,\mbox{km}\,\mbox{s}^{-1}$ at the solar radius $R_{\mbox{\sun}}=8$\,kpc as shown in the top panel of Fig.\,\ref{fig:adiabatic}. 

We apply slight perturbation on circular orbits that mimic the radial velocity dispersion of $40\,\mbox{km}\,\mbox{s}^{-1}$ at the solar Galactocentric radius, radial-to-vertical velocity dispersion ratio of $\sqrt{2}$ \citep{fuc09} and velocity scalelength of $3.5$\,kpc (BO12). We define the variation of action $\Delta_{J}$ to be $1\sigma$ over the mean value of the action for each time step. The results of $\Delta_{J}$ are listed in Table\,\ref{table:adiabatic}. The results show that the approximated actions are almost conserved in all cases. 

We also examine the potential due to a double-exponential disc (i.e. with density decaying both radially and vertically) \citep[see appendix A, equation A10,][]{kui89}, and also the flattened isothermal potential:
\begin{eqnarray}
\label{eq:isothermal_potential}
\Phi=V^2_c(R_{\mbox{\sun}})\ln\sqrt{R^2+\frac{z^2}{q^2}}.
\end{eqnarray}

\noindent
The results are qualitatively similar, with the variation of actions less than $5$ per cent in all cases. \citet{bin12} studies a more precise method of calculating the actions, but it is computationally more demanding. As we will discuss in Section\,\ref{sec:constrain_potential}, we find that the error in using the adiabatic approximation will introduce a systematic error of about $1$\,per cent in estimating the potential parameters. The adiabatic approximation is sufficient for our current study since the purpose of this study is to introduce a new statistical method to constrain the Galactic potential.

We only focus on the Miyamoto--Nagai potential and the flattened isothermal potential in this study. We do not consider the potential due to a double-exponential disc because the analytic form of this potential (by solving the Poisson equation) involves an improper integration of a Bessel function \citep{kui89}. This is computationally very expensive. Moreover, the Miyamoto--Nagai potential is good enough as a first approximation for a realistic disc potential.

%
%
%
%
%
%
\section[]{Matching the phase-space distribution of mono-abundance populations}\label{sec:DF-match}
In this section, we will show that the DF in equation\,(\ref{eq:DF}) is a good representation of the observed position--velocity distribution of mono-abundance subpopulations of the Milky Way disc, namely that they exhibit an exponential spatial density, both radially and vertically, and an isothermal velocity dispersion in the vertical direction and exponentially decaying in the radial direction. 
\begin{figure*}
\begin{center} 
$\begin{array}{cc}
\includegraphics[scale=0.27]{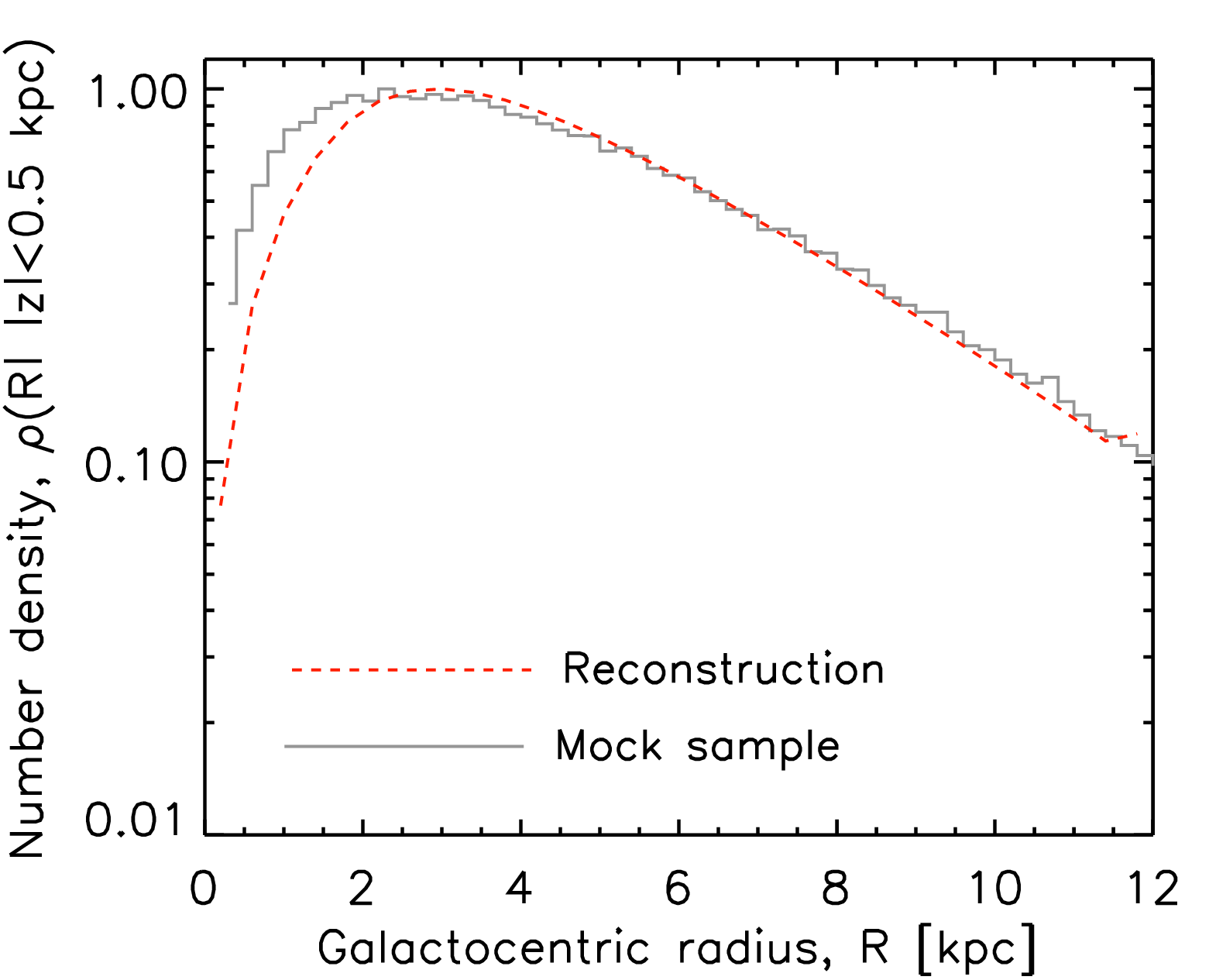} & 
\includegraphics[scale=0.27]{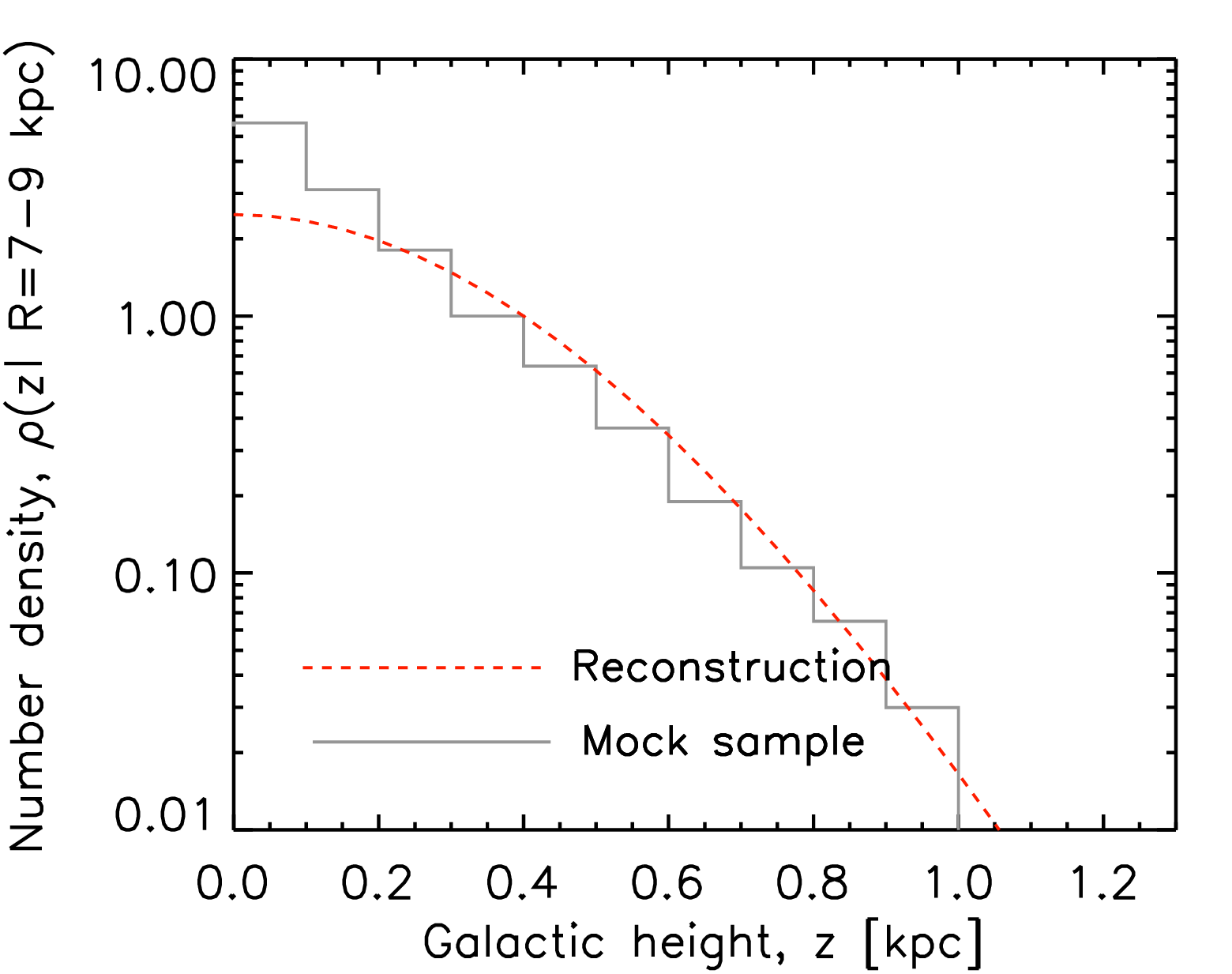} 
\end{array}
$\\
$\begin{array}{ccc}
\includegraphics[scale=0.23]{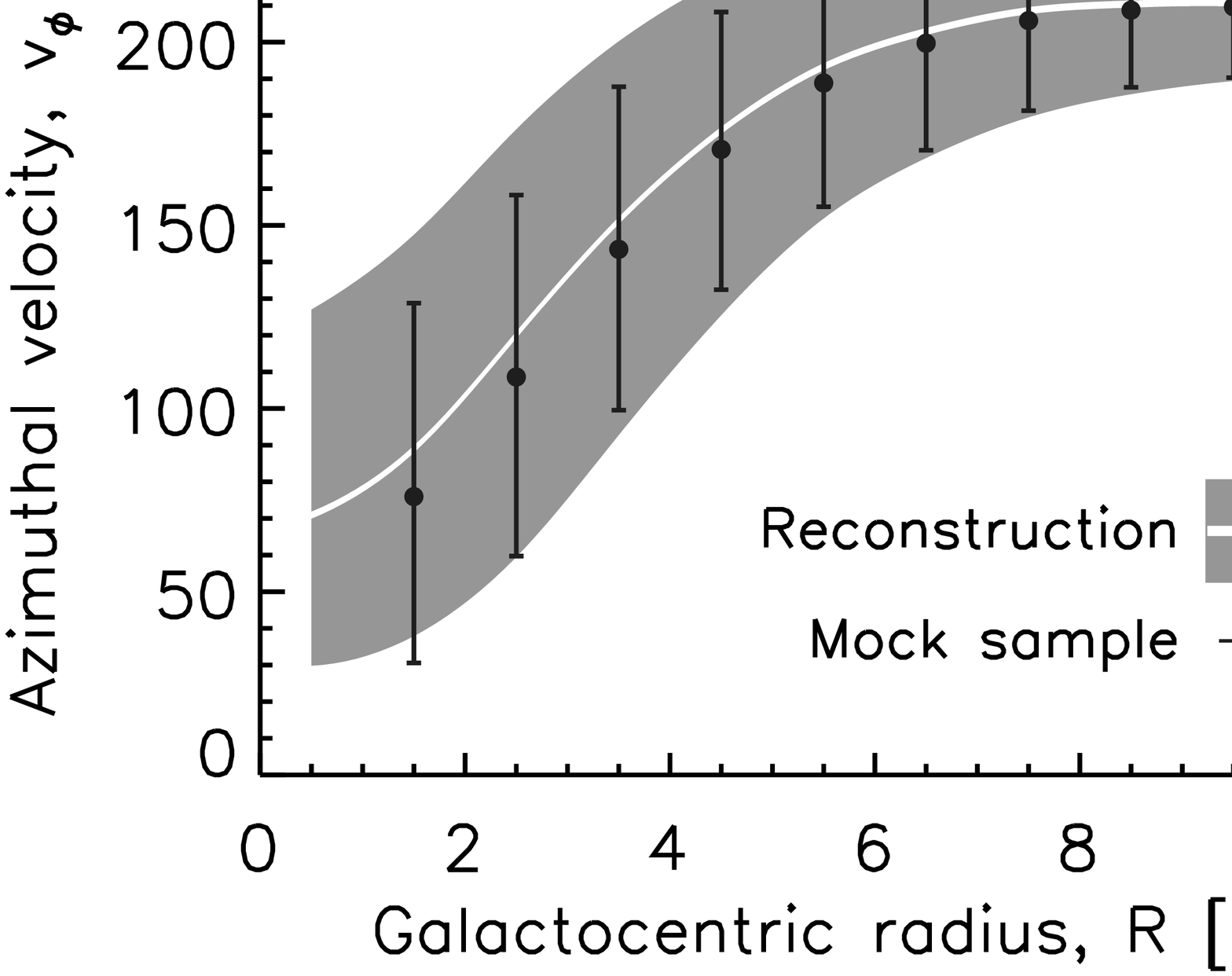} & \hspace{0.2cm}
\includegraphics[scale=0.23]{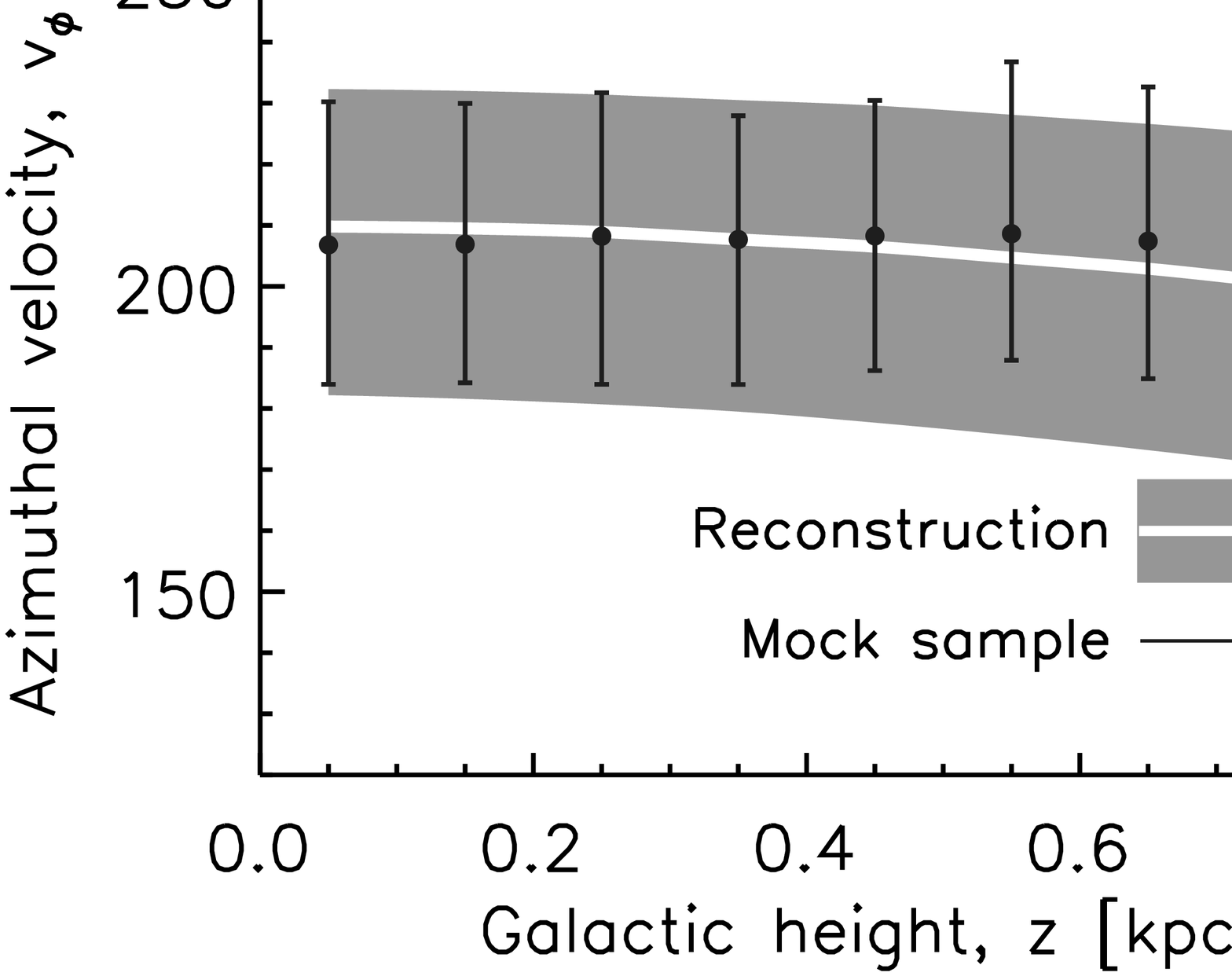} & \hspace{0.2cm}
\includegraphics[scale=0.23]{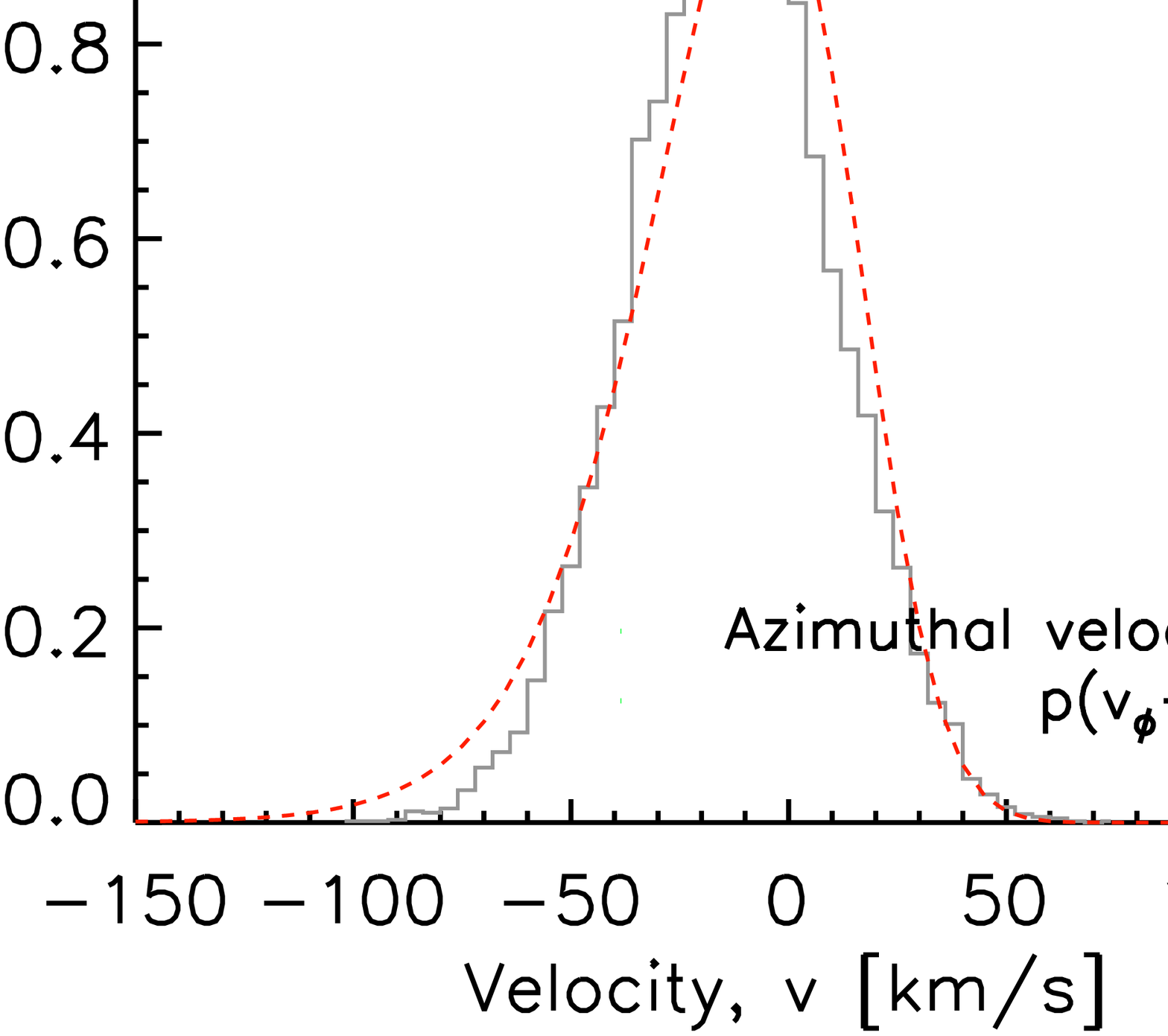} 
\end{array}
$
\vspace{0.5cm}
\caption{How well can the DF family from equation\,(\ref{eq:DF}) approximate the observed properties of Galactic mono-abundance stellar subpopulations (BO12) in a Miyamoto--Nagai potential? Those mono-abundance populations are observed to have an approximate exponential spatial density, both radially and vertically, and an isothermal velocity dispersion in the vertical direction and exponentially decaying in the radial direction. To create mock data, we assume a radial velocity dispersion of about $40\,\mbox{km}\,\mbox{s}^{-1}$ at the solar Galactocentric radius and scaleheight $h_z=0.18$\,kpc for the mock data. The top row shows the radial (left) and vertical (right) density distributions in $(R,z)\in[0,12]\times[-0.5,0.5]\,\mbox{kpc}^2$ and $(R,z)\in[7,9]\times[0,1.3]\,\mbox{kpc}^2$, respectively. The dashed red line shows the reconstructed profile from the best-fitting DF and the grey solid histogram shows the mock data. The second row shows the distribution of $v_R$ as a function of $R$ with $z\in[-0.5,0.5]$\,kpc (left-hand panel) and as a function of $z$ with $R\in[7,9]$\,kpc (middle panel), and also the histogram in $(R,z)\in[7,9]\times[-0.5,0.5]\,\mbox{kpc}^2$. The grey shaded region shows the reconstructed profile from the best-fitting DF with the white solid line to be the median value, whereas the black solid filled circle and their vertical error bars show the results of the mock data. The third and fourth rows show the distribution of $v_z$ and $v_\phi$, respectively. The DF family fits the tracer population very well for a disc potential.}
\label{fig:NM_fit}
\end{center}
\end{figure*}

%
%
%
%
%
%
\subsection[]{Dealing with the Jacobian change of measure}\label{sec:jacobian}
In order to have the conversion of DF between canonical position--velocity variable and the action variable, one has to take care of the Jacobian term carefully. Note that
\begin{eqnarray}
\int\dd^3{\boldsymbol x}\,\dd^3{\boldsymbol p}\,\widetilde{f}({\boldsymbol x},{\boldsymbol p})=\int\dd^3{\boldsymbol J}\,\dd^3{\boldsymbol \theta}\,f({\boldsymbol J}).
\end{eqnarray}

\noindent
where ${\boldsymbol p}$ is the canonical momentum associated with ${\boldsymbol x}$. Since $({\boldsymbol x},{\boldsymbol p})$, $({\boldsymbol J},{\boldsymbol \theta})$ are canonical variables and measure is invariance under canonical transformation, we have
\begin{eqnarray}
\widetilde{f}({\boldsymbol x},{\boldsymbol p})=f\big({\boldsymbol J}({\boldsymbol x},{\boldsymbol p})).
\end{eqnarray}

Furthermore, in a cylindrical coordinate, $(p_R,p_\phi,p_z)=(v_R , R v_\phi, v_z)$, we deduce
\begin{eqnarray}
f({\boldsymbol x},{\boldsymbol v})=Rf({\boldsymbol J}({\boldsymbol x},{\boldsymbol p}({\boldsymbol x},{\boldsymbol v}))).
\label{eq:configuration_space}
\end{eqnarray}

In summary, for an axisymmetric potential, in order to convert an action-based DF to $({\boldsymbol x},{\boldsymbol v})$ space distribution, it suffices to multiply the Jacobian ($=R$) term.

%
%
%
%
%
%
\subsection{Fitting the phase-space distribution via maximum likelihood}
We create mock data stellar tracers with $(R,z,v_R,v_\phi,v_z)$ attributes, such that it has a density scalelength of about $2.5$\,kpc, dispersion scalelength of about $3.5$\,kpc with a Gaussian velocity dispersion of $\sigma_R=40\,\mbox{km}\,\mbox{s}^{-1}$ at the solar Galactocentric radius, $\sigma_R/\sigma_z=\sqrt{2}$, $\sigma_R/\sigma_\phi=1.5$ \citep[][BO12]{fuc09}. We assume that the mean velocity of the Gaussian for each spatial point is
\begin{eqnarray}
\overline{v_R}(R,z)=\overline{v_z}(R,z)=0
\end{eqnarray}
\begin{eqnarray}
\overline{v_\phi}(R,z)=\overline{v_\phi}(R)=V_c(R)-A\frac{\sigma_R^2(R)}{2V_c(R)},
\end{eqnarray}

\noindent
where (see \citealt{bov12} for details)
\begin{eqnarray}
A=\bigg(\frac{1}{h_R}+\frac{1}{h_\sigma}\bigg)\times R-0.5.
\end{eqnarray}

We assume that all of the mixed moments ($\sigma^2_{R\phi}$, $\sigma^2_{Rz}$ and $\sigma^2_{\phi z}$) vanish. Thus, we assume that both the vertex deviation and the tilt of the velocity ellipsoid are zero, even though these were not measured by BO12. The tilt of the velocity ellipsoid in particular is not expected to be zero at heights $\gtrsim1$\,kpc above the plane \citep[e.g.][]{sib08}. However, as discussed by \citet{bin11b}, the adiabatic approximation is unable to capture a non-zero tilt as it assumes that the radial and vertical motions are independent. Thus, even though the quasi-isothermal DF of equation\,(\ref{eq:DF}) has a non-zero tilt when used with correctly calculated actions \citep[for example, using the torus machinery;][]{bin11b}, it does not when using the adiabatic approximation. For this reason, we assume that the tilt of the velocity ellipsoid is zero in what follows. 
\begin{figure}
\begin{center} 
\includegraphics[scale=0.3]{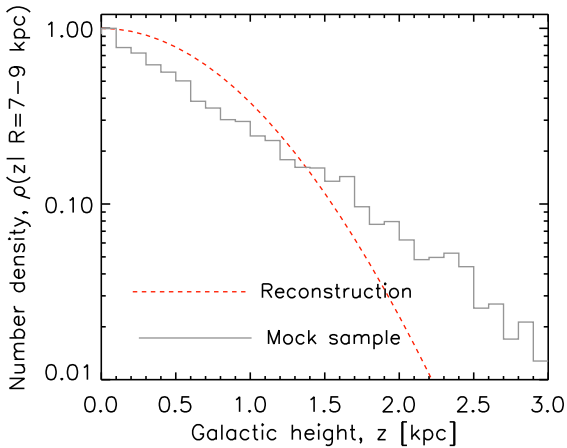} 
\caption{Similar to the top-right panel of Fig.\,\ref{fig:NM_fit}, but for the isothermal potential with the axis ratio $q=1$ and circular velocity $V_c=220\,\mbox{km}\,\mbox{s}^{-1}$. We also tried various choices of $q$, but the fits are equally unsatisfactory: the DF family in equation\,(\ref{eq:DF}) cannot produce a vertically exponential tracer density profile, if stars orbit in a flattened isothermal sphere.}
\label{fig:log_fit}
\end{center}
\end{figure}

We search for the best-fitting DF parameters by maximum likelihood. For this section, we fix a particular choice of Miyamoto--Nagai potential parameters: $GM=7.5\times 10^5\,\mbox{km}^2\,\mbox{s}^{-2}\,\mbox{kpc}^{-1}$, $a=5$\,kpc, $b=1$\,kpc. We examine various choices of the potential parameters. It does not change the results qualitatively. We emphasize that the mock tracer population mimics a mono-abundance population. Note that the probability that we observe $({\boldsymbol x}_\alpha,{\boldsymbol v}_\alpha)$, according to our DF model with DF parameters ${\boldsymbol \lambda}\df$, is
\begin{eqnarray}
\mbox{Pr}({\boldsymbol x}_\alpha,{\boldsymbol v}_\alpha|\Phi,{\boldsymbol \lambda}\df)=f({\boldsymbol x}_\alpha,{\boldsymbol v}_\alpha|\Phi,{\boldsymbol \lambda}\df).
\end{eqnarray}

Given this probability, we can define the log likelihood to be the sum over $N_\alpha$ mock data, and search for the best DF parameters by maximizing the log likelihood,
\begin{eqnarray}
\ln\mathcal{L}\df({\boldsymbol \lambda}\df)\equiv\sum_{\alpha}\ln[\mbox{Pr}({\boldsymbol x}_\alpha,{\boldsymbol v}_\alpha|\Phi,{\boldsymbol \lambda}\df)].
\end{eqnarray}

We obtain the best-fitting parameters by performing a nested-grid search on the multidimensional ${\boldsymbol \lambda}\df$ space and finding the set of parameters that maximizes the likelihood. After obtaining the best-fitting parameters, we calculate from $f({\boldsymbol J}({\boldsymbol x},{\boldsymbol v}))$ the stellar profile in configuration space $({\boldsymbol x},{\boldsymbol v})$ and compare it to the one of the mock data to check whether our assumption on the functional form of the DF represents well the mock data. We consider a grid of position--velocity configuration space $(R,z,v_R,v_\phi,v_z)$ and calculate the DF in configuration space using equation\,(\ref{eq:configuration_space}). We then calculate the stellar profile of each position--velocity component by marginalizing over the other spatial-velocity components.

Note that, given the functional form of the DF in equation\,(\ref{eq:DF}), the predicted vertical profile is fixed by the vertical velocity dispersion and the potential, but independent of the scaleheight of the mock data. We find that the vertical profile is predicted satisfactorily with the Miyamoto--Nagai potential (as shown in Fig.\,\ref{fig:NM_fit}) but not with the flattened isothermal potential (as shown in Fig.\,\ref{fig:log_fit}). 

This can be explained by the following. In the case where $R\gg z$, the vertical restoring force of the Miyamoto--Nagai potential and thus $\dot{z}$ is about constant. This implies that the vertical action is proportional to $z$, and the vertical profile in ansatz equation\,(\ref{eq:DF}) goes down exponentially. On the other hand, the vertical restoring force of the flattened isothermal potential and thus $\dot{z}$ is proportional to $z$. This implies that the vertical action is proportional to $z^2$, and the vertical profile will go down too drastically [proportional to $\exp(-z^2)$]. 

Also note that, in the upper panel of Fig.\,\ref{fig:NM_fit}, the DF predicts a shallower vertical density profile near the Galactic plane. As SEGUE does not observe the Galactic plane, we cannot decide whether this prediction is accurate or one should revise the DF to better fit the vertical density profile at this point. Nonetheless, the density distribution is not expected to be exponential all the way to $z=0$ and $R=0$. Our aim here is to match the exponential profiles in the range of $R$ and $z$ where SEGUE has data.

We conclude that the quasi-isothermal DF of equation\,(\ref{eq:DF}) provides a good representation of the DF of mono-abundance subpopulations of the Milky Way disc, namely that the DF predicts (1) an exponential spatial density, both radially and vertically; (2) an isothermal velocity dispersion in the vertical direction and (3) exponentially decaying in the radial direction. More importantly, the action-based DFs predict a low-$v_\phi$ tail. As discussed in Section\,\ref{sec:introduction}, this coincides better with the observation \citep[e.g.][]{fuc09}. In the following section, we will generate mock samples by rejection sampling {\em directly from the DF}.

%
%
%
%
%
%
\section[]{Constraining the Galactic potential}\label{sec:constrain_potential}
\subsection[]{Spatial selection function}
In this section, we will show that, given a functional form of the potential and the DF, we can recover the optimal potential parameters by marginalizing over the DF parameters. Note that in practice, we only observe a very small part of the Galaxy. In this study, we assume that the observed volume is a cylinder around the Sun with a finite width and height. We assume that the location of the Sun in the Galactocentric Cartesian coordinate is $(x_{\mbox{\sun}},y_{\mbox{\sun}},z_{\mbox{\sun}})=(8\,\mbox{kpc},0\,\mbox{kpc},0\,\mbox{kpc})$ and include the spatial selection function
\begin{eqnarray}
P\select=\left\{
\begin{array}{l}
1,\,\,\mbox{if}\,\sqrt{(x-x_{\mbox{\sun}})^2+y^2}<1\,\mbox{kpc},|z|<2.5\,\mbox{kpc}\\
0,\,\,\mbox{otherwise}.
\end{array}
\right.
\label{Eq:p-select}
\end{eqnarray}

With this selection function in place, given any parameters of the DF and the potential, ${\boldsymbol \lambda}\equiv({\boldsymbol \lambda}\df,{\boldsymbol \lambda}_{\mbox{\tiny potential}})$, we carefully normalize the DF through Monte Carlo integration, more precisely
\begin{eqnarray}
f({\boldsymbol J}|{\boldsymbol \lambda})\longrightarrow\frac{f({\boldsymbol J}|{\boldsymbol \lambda})}{\mathcal{N}({\boldsymbol \lambda})},
\end{eqnarray}

\noindent
where 
\begin{eqnarray}
\mathcal{N}({\boldsymbol \lambda})=\frac{1}{2\upi}\int\dd^3{\boldsymbol x}\,\dd^3{\boldsymbol v}\;R\,P\select(R,\phi,z)f({\boldsymbol J}({\boldsymbol x},{\boldsymbol v})|{\boldsymbol \lambda}).
\end{eqnarray}

To our best knowledge, this is the first attempt that has never before been implemented to study the joint distribution of DF and potential parameters in the context of constraining the Galactic potential. 

\subsection[]{In-plane Miyamoto--Nagai potential}
First, we consider the two-dimensional in-plane Miyamoto--Nagai potential which contains two potential parameters: $GM$ and $a$. We consider the case where $h_R=2.5$\,kpc, $h_\sigma=3.5$\,kpc and $\sigma^2_{R}=110\,\mbox{km}\,\mbox{s}^{-1}$ in equation\,(\ref{eq:DF}), which roughly mimics the observed velocity distributions from our calculation in Section\,\ref{sec:DF-match}. 
\begin{figure*}
\begin{center} 
\includegraphics[scale=0.42]{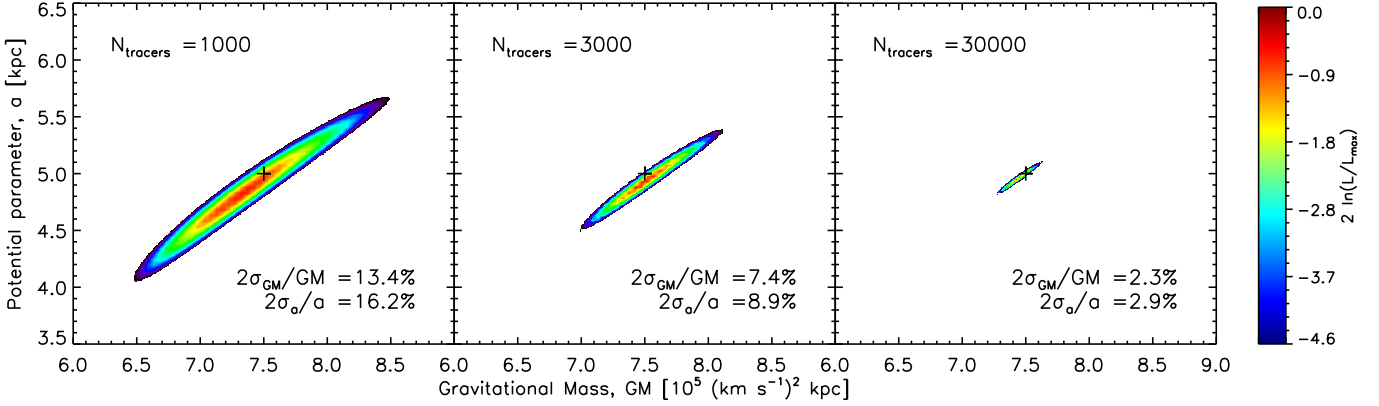} 
\caption{Likelihoods of the data given the parameters of a Miyamoto--Nagai potential, assuming that the $N_{\mbox{\tiny tracers}}$ are confined to a cylinder of radius $1$\,kpc around the Sun up to $|z|<2.5$\,kpc. The black cross symbols indicate the true value that we use to generate the mock data. We only plot the likelihood region that encloses $95$\,per cent of the total probability. The symmetrized uncertainty corresponding to $95$\,per cent significance (two-sigma) of each parameter over the true value is noted at the bottom right of each panel. The left-hand panel shows the results of mock data points $N_{\mbox{\tiny tracers}}=1000$, the middle panel $N_{\mbox{\tiny tracers}}=3000$ and the right-hand panel $N_{\mbox{\tiny tracers}}=30\,000$: the likelihood contours tighten as $1/\sqrt{N_{\mbox{\tiny tracers}}}$ around the correct input parameters for the mock data. The data constrain a highly degenerate combination of $GM$ and $a$. The uncertainties are larger than the one in the study of a flattened isothermal potential because the parameters are highly correlated.}
\label{fig:NM_2D_opti_potential}
\end{center}
\end{figure*}
\begin{figure*}
\begin{center} 
\includegraphics[scale=0.42]{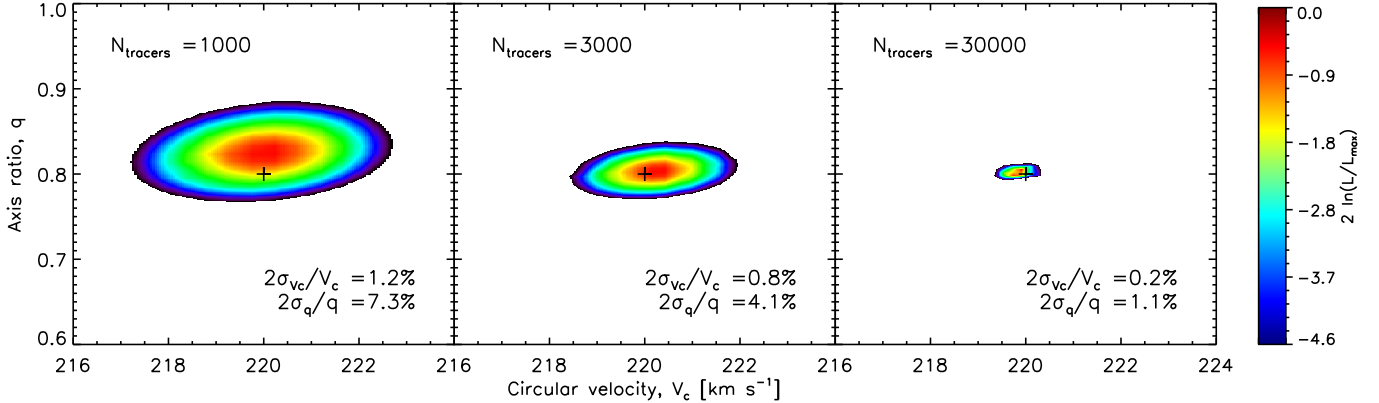} 
\caption{Similar to Fig.\,\ref{fig:NM_2D_opti_potential} but for a flattened isothermal potential. Here the estimates for the potential flattening and the circular velocity are hardly correlated.}
\label{fig:log_3D_opti_potential}
\end{center}
\end{figure*}

We define the likelihood of the potential parameters, $\mathcal{L}=\mathcal{L}({\boldsymbol \lambda}_{\mbox{\tiny potential}})$, by marginalizing over ${\boldsymbol \lambda}\df$. We estimate the one-sigma significance of the best-fitting potential parameter by finding the demarcation line at $2\ln(\mathcal{L}/\mathcal{L}_{\max})=2.7$ and two-sigma significance at $2\ln(\mathcal{L}/\mathcal{L}_{\max})=4.6$. We create data with $GM=7.5\times10^5\,\mbox{km}^2\,\mbox{s}^{-2}\,\mbox{kpc}^{-1}$, $a=5$\,kpc, $b=1$\,kpc in the observed volume $1$\,kpc radially from the Sun and $|z|<2.5$\,kpc.

We reiterate the process for various sample sizes for the mock data, $N_{\mbox{\tiny tracers}}$ for the same observed volume. As shown in Fig.\,\ref{fig:NM_2D_opti_potential}, the uncertainty goes down as $1/\sqrt{N_{\mbox{\tiny tracers}}}$. We use the results with the sample size $N_{\mbox{\tiny tracers}}=30\,000$ to be our reference, because the uncertainty of the potential parameters is $\leq 3$\,per cent in this case. We show that this reference value coincides with the true value, indicating that, assuming a potential model, we can recover the true potential parameter with observation restricted to a small volume. We show that we can find the best-fitting potential parameters within one- to two-sigma significance from the true value regardless of the sample size. 

More importantly, the data constrain a highly degenerate combination of $GM$ and $a$: since $V_c^2 \propto M/a$, we find that $M \propto a$ as seen in Fig.\,\ref{fig:NM_2D_opti_potential}. On top of that, we show that, with the sample size of $3000$, we recover both parameters with a precision of about $8$\,per cent for a $95$\,per cent significance (two-sigma). The uncertainties are larger than the one in the study of a flattened isothermal potential below because the parameters are highly correlated.

\subsection[]{Flattened isothermal potential}
We extend this result to the three-dimensional case by considering the flattened isothermal potential with two parameters (equation\,\ref{eq:isothermal_potential}): $V_c$ and $q$. We choose to constrain only two potential parameters because (1) the main purpose of this study is to show that we can pin down the two upmost important parameters of the Galactic potential around the solar neighbourhood, namely the total mass and the vertical density scaleheight. In this regard, two parameters are sufficient. (2) While our method is general and could be used to constrain any number of potential parameters, it is computationally expensive to add another dimension in the nested-grid optimization method that we use. We could have implemented Markov chain Monte Carlo (MCMC) and have increased the efficiency of higher dimensional search, but the purpose of this paper is to illustrate the new method. We shall leave the MCMC implementation to a later study.

We consider the warm-disc case where $h_R=2.5$\,kpc, $h_\sigma=3.5$\,kpc, $\sigma^2_{R}=110\,\mbox{km}\,\mbox{s}^{-1}$ and $\sigma^2_{z}=90\,\mbox{km}\,\mbox{s}^{-1}$ for DF in equation\,(\ref{eq:DF}). We create the mock data from the potential with $V_c=220\,\mbox{km}\,\mbox{s}^{-1}$ and $q=0.8$. As shown in Fig.\,\ref{fig:log_3D_opti_potential}, one can recover both the circular velocity $V_c$ and the axis ratio $q$ successfully with our proposed method. 

\begin{figure}
\begin{center} 
\includegraphics[scale=0.32]{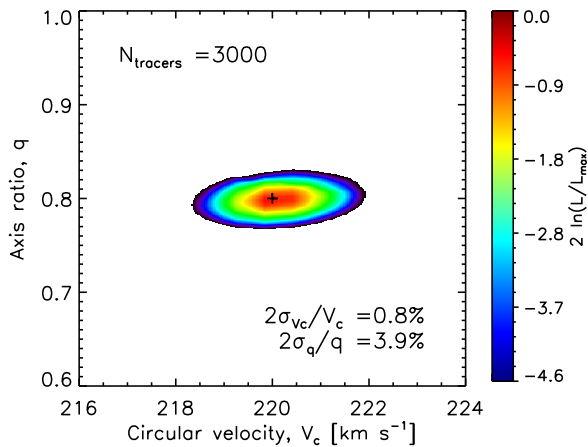} 
\caption{Similar to the middle panel of Fig.\,\ref{fig:log_3D_opti_potential}, but with the vertical range of available tracer constraints restricted to $1<z<2.5$\,kpc -- the mid-plane is cutout: this is a more realistic selection function since in most cases, without a Galactic dust map, observations in the equatorial Galactic disc are less reliable.}
\label{fig:log_3D_opti_potential_without_midplane}
\end{center}
\end{figure}

Note that, in most cases, without a Galactic dust map, observations in the equatorial Galactic disc are less reliable. Therefore, we also tested the more realistic scenario, where the observed volume is restricted to the cylindrical height of $1<z <2.5$\,kpc without the mid-plane. The result as shown in Fig.\,\ref{fig:log_3D_opti_potential_without_midplane} shows that we can about equally well constrain the axis ratio $q$ and the circular velocity $V_c$, given the same sample size. This could be probably due to the fact that our three-dimensional flattened isothermal potential is too restrictive; therefore, the axis ratio is not too sensitive to the vertical observed volume restriction. In both cases, we show that, with the sample size of $3000$, we recover the circular velocity with a precision of about $1$\,per cent for a $95$\,per cent significance (two-sigma) and of about $4$\,per cent for the axis ratio.

\subsection[]{Marginalizing versus maximizing over the DF parameters}

It is important to note that, in this study, we perform the potential parameter estimation by marginalizing over the DF parameters instead of maximizing over the DF parameters. It has been argued that this is the best way to constrain the potential in this type of study \citep{mag06,mag13}. To the best of our knowledge, our study is also the first successful study in implementing this marginalization.

Interestingly, we find that by performing marginalization, the results we obtain are almost exactly the same as maximizing over the DF parameters. Since marginalization will take at least an order of magnitude more in terms of computational time, it might be worthwhile to explain these results. To illustrate these results, we consider a two-dimensional parameter space $(x,y)$ and we marginalize over one of the dimensions. It is trivial to generalize the arguments to higher dimensions.

If the posterior joint distribution is a two-dimensional joint Gaussian probability distribution, assuming flat prior, the posterior probability distribution is given by
\begin{eqnarray}
P(x,y|\mbox{data})\propto\exp\bigg(-\frac{\chi_R^2}{2}\bigg).
\end{eqnarray}

Since this is a monotonic function, the contour of the posterior probability distribution will resemble the contour of $\chi_R^2$. Note that the general formula for a two-dimensional posterior Gaussian probability distribution can be written as
\begin{eqnarray}
P(x,y)\propto\exp\Bigg[-\frac{1}{2(1-\rho^2)}\bigg(\frac{x^2}{\sigma_x^2}+\frac{y^2}{\sigma_y^2}-\frac{2xy\rho}{\sigma_x\sigma_y}\bigg)\Bigg].
\end{eqnarray}

One can show that the maximum of $P(y|x)$ is obtained at $y_{\mbox{\tiny max}} = \frac{\rho \sigma_y}{\sigma_x}$. Therefore,
\begin{eqnarray}
P(x,y_{\mbox{\tiny max}})\propto\exp\bigg(-\frac{x^2}{2\sigma_x^2}\bigg).
\end{eqnarray}

Differing only by a multiplicative constant, this is the same as 
\begin{eqnarray}
P(x)=\int P(x,y)\dd y,
\end{eqnarray}

When considering the difference in $\chi_R^2$ with respect to the minimum point, the multiplicative constant will drop out. Hence, with all the arguments above, we show that in the marginalized $P(x)$, to obtain the boundary point at which $\chi_R^2 \rightarrow\chi_{R,\mbox{\tiny min}}^2+\frac{1}{N-2}$, it is the same as maximizing over parameter $y$, with the assumption that the posterior is a multidimensional Gaussian distribution.

\subsection[]{Sources of uncertainties}
To better address the concern of systematic error arising from the use of the adiabatic approximation in the potential parameter estimation, we also perform an identical study with mock data generated using the torus machinery (McMillan \& Binney, private communication). The torus machinery \citep{bin11b}, compared to the adiabatic approximation, is an more accurate method of calculating actions. We find that the best-fitting parameters (recovered assuming the adiabatic approximation), in this case, are slightly shifted (about $1$\,per cent) compared to the input parameters (generated using the torus machinery). This is probably due to the error in calculating actions with the adiabatic approximation. We conclude that more accurate and efficient ways of calculating actions are very valuable in improving our analysis; however, our analysis with the adiabatic approximation only introduces a non-significant error.

We also test our method by shifting the Galactocentric frame velocities by $10\,\mbox{km}\,\mbox{s}^{-1}$, mimicking the uncertainty due to the solar motion. We find that all the results presented in this paper remain to be qualitatively similar, with the only difference that the estimated circular velocity of the Galactic potential will be shifted by the same amount. More importantly, when the Galactocentric frame is shifted by a larger amount, the maximum likelihood decreases more, accordingly. Therefore, if we consider the solar motion to be another free parameter in the fit, we could study the most probable solar motion with our proposed method and put an independent constraint on the true solar motion in the Galactocentric frame.

\subsection[]{Kinematically cold versus warm populations}
BO12 show that the Galactic disc, when decomposed into mono-abundance populations, contains cold populations (i.e. smaller velocity dispersion) as well as warmer populations. To compare the constraints which we can obtain from different populations, we study the three-dimensional Miyamoto--Nagai potential. We consider a warm population with $\sigma^2_{R}=110\,\mbox{km}\,\mbox{s}^{-1}$, $\sigma^2_{z}=90\,\mbox{km}\,\mbox{s}^{-1}$ as before, and a cold population with $\sigma^2_{R}=75\,\mbox{km}\,\mbox{s}^{-1}$, $\sigma^2_{z}=60\,\mbox{km}\,\mbox{s}^{-1}$. We fit for $b$ (the scaleheight) and $GM$ (the total mass), but fix $a$ (the scalelength) to keep the computational cost down (see above). We find that cold populations constrain the total mass $GM$ better and also constrain slightly better the scaleheight $b$. The result is expected for $GM$ because it is related to the circular velocity. One should be able to pin down the circular velocity better if the dispersion of the velocity is smaller. We interpret the result for the scaleheight as follows: there are two competing effects for the scaleheight estimation: (1) the warm populations probe a wider range in height, given the same selection function; (2) the cold populations can pin down the trajectories/orbits better. We find that in our case, the latter effect slightly dominates the former. However, the results should be viewed with caution. The results could be biased because (1) our model might be too restrictive, (2) the selection function that we choose ($|z|< 2.5$\,kpc) is larger than the Galactic scaleheight; therefore, there is not much gain in probing a wider range in height. With these caveats in mind, we find that colder populations are more valuable in the study of the Galactic potential with our proposed method.

%
%
%
%
%
%
\section[]{Conclusions}\label{sec:conclusion}
We demonstrate a way to constrain the Galactic potential from mono-abundance stellar populations. We show that the phase-space distribution properties of mono-abundance stars can be fitted very well with a simple action-based DF. Assuming that the vertical and radial velocity dispersion scalelengths are the same, these properties are fully determined by only four DF parameters. 

We further show that, assuming that the proposed DF is the true representation of the mono-abundance populations and certain parametrization of the Galactic potential, we devise a statistical and rigorous method to measure the Galactic potential parameters. This is achieved by calculating the likelihood of observational data, given a joint set of parameters for both the DF and the gravitational potential and subsequently marginalizing over the DF parameters. \citet{mcm13} perform a rigorous test on the method we propose in this paper and confirm that our method has significant advantages on both the uncertainty in the potential parameter estimation and the computational time.

We create a mock data sample from the DF with various potentials, including the Miyamoto--Nagai in-plane potential and a flattened isothermal potential. We show that even with a mono-abundance sample of the size of a few thousands, given a two-parameter gravitational potential, we can pin down the potential parameters to a few per cent. 

In this study, we assume that the measurement is perfect without uncertainty in the position and velocity measurements. The inclusion of measurement uncertainty should be conceptually straightforward.

Although a more realistic potential, for instance a combination of halo, bulge and disc potential, will contain more than two parameters and hence post a challenge to the estimation, different mono-abundance stars are tracing the same Galactic potential and therefore their joint constraints with the method we propose are still very promising. In the current available SEGUE survey, considering an elemental abundance bin of [Fe/H]$=0.1$ and $[\alpha/\mbox{Fe}]=0.05$, we have about $100$--$1000$ tracers per bin. In the future, when GALAH is fully operational, we expect to have about $20\,000$ tracers per bin. In practice, when dealing with multiple mono-abundance populations, one can combine the constraints from different populations by multiplying the individual likelihoods marginalized over the DF parameters.

The assumptions of axisymmetry, of DFs that are smooth in the actions and no time variation remain important limitations in our current study. To tackle some of these aspects, we have started analysing data from the large $N$-body simulation described in \citet*{don13}, which has strong, time-dependent spiral structure. As a preliminary result, we find that, even with realistic levels of non-axisymmetry and time variation of the potential, the method still works, but we plan to describe this in a later paper for a more concrete and rigorous analysis of this problem.

\section*{Acknowledgements}
We thank the anonymous referee for helpful comments. We also thank Elena D'Onghia for providing data from the simulation in \citet{don13}. YST and JB are grateful to the Max-Planck-Institut f\"{u}r Astronomie for its hospitality and financial support during the period in which this research was performed. JB was supported by NASA through the Hubble Fellowship grant HST-HF-51285.01 from the Space Telescope Science Institute, which is operated by the Association of Universities for Research in Astronomy, Incorporated, under the NASA contract NAS5-26555. JB was also partially supported by SFB 881 funded by the German Research Foundation DFG.

\appendix

\label{lastpage}

\end{document}